\documentstyle[psfig,conf_iap]{article}
\def\lesssim{\mathrel{\hbox{\rlap{\hbox{\lower4pt\hbox{$\sim$}}}\hbox{$<$}}}}
\def\gtrsim{\mathrel{\hbox{\rlap{\hbox{\lower4pt\hbox{$\sim$}}}\hbox{$>$}}}}
\begin{document}
\heading{%
%
Weak Lensing by Large-Scale Structure with the FIRST
Radio Survey
%
} 
\par\medskip\noindent
\author{%
A. Refregier$^{1,2}$, S. T. Brown$^{2}$, M. Kamionkowski$^{2}$,
D. J. Helfand$^{2}$, C. M. Cress$^{2}$, A. Babul$^{3}$, R. Becker$^{4}$,
R. L. White$^{5}$
}
\address{Department of Astrophysical Sciences, Peyton Hall,
Princeton University, Princeton, NJ 08544}
\address{Columbia Astrophysics Laboratory, 538 W. 120th Street, New York,
NY 10027}
\address{Department of Physics \& Astronomy, University of Victoria, P.O. Box
3055, Victoria, BC V8W sP6, Canada}
\address{Institute of Geophysics and Planetary Physics, Lawrence
Livermore National Laboratory, Livermore, CA 94450}
\address{Space Telescope Science Institute, 3700 San Martin Drive,
Baltimore, MD 21218}

\begin{abstract}
The coherent image distortions induced by weak gravitational lensing
can be used to measure the power spectrum of density inhomogeneities
in the universe. We present our on-going effort to detect this effect
with the FIRST radio survey, which currently contains about 400,000
sources over 4,200 square degrees, and thus provides a unique
resource for this purpose. We discuss the sensitivity of our
measurement in the context of various cosmological models. We then
discuss the crucial issue of systematic effects, the most serious of
which are source fragmentation, image-noise correlation, and VLA-beam
anisotropy. After accounting for these effects, we expect our
experiment to yield a detection, or at least a tight upper limit, for
the weak lensing power spectrum on 0.2-20 degree scales.
\end{abstract}

\section{Introduction}
The coherent image distortions induced by weak gravitational lensing
can be used to measure the power spectrum of density inhomogeneities
in the universe (see \cite{nar96} for a review, and \cite{ref98a} for
a bibliography). While most searches for this effect are conducted in
the optical band, we are engaged in an effort to detect this effect
with the FIRST radio survey. The main advantages of our experiment are
the high redshift of the sources ($\langle z \rangle \sim 1$), the
wide solid angle covered by the FIRST survey (currently 4,200
deg$^{2}$), and the reproducibility of the systematic effects. While a
detailed account of our study will appear in ref.~\cite{ref98c} (see
also \cite{ref98a}), its theoretical framework and the study of one of
the systematic effects are presented in refs.~\cite{kam98} and
\cite{ref98b}.

\section{Survey and Methods}
The FIRST radio survey \cite{bec95,whi97} was conducted at the VLA
at 1.4 GHz in the B configuration. Its $5\sigma$ flux limit is
0.75 mJy, with a restoring beam FWHM of $5''.4$ and a pixel size of
$1''.4$. The survey currently contains about $4\times 10^5$ sources
over $A \simeq 4,200$ deg$^2$, with a mean redshift of $\langle z \rangle
\sim 1$. Observing time has been allocated to extend
its coverage to 7,200 deg$^2$, while its nominal area is 10,000 deg$^{2}$.

We characterize each source by its ellipticity, $\epsilon_{i} \equiv
\frac{a^2-b^2}{a^2+b^2} \{\cos 2\alpha,\sin 2\alpha\}$, where $a$ and
$b$ are the deconvolved major and minor axes, and $\alpha$ is the
position angle, derived from elliptical gaussian fits. For our
weak-lensing search, we only keep resolved sources ($a>2^{\prime
\prime}$, $b>0$), which represent 38\% of the total number of
sources and amount to a source density of $n\simeq 36$ deg$^{-2}$.

The weak lensing shear $\gamma_{i}$ is related to the source-averaged
ellipticity by $\langle \epsilon_{i} \rangle \simeq -g \gamma_{i}$,
where $g\equiv 2(1-\sigma_{\epsilon}^{2})$ is the shear-ellipticity
conversion factor, and $\sigma_{\epsilon}^{2} = \langle
\epsilon_{1}^{2} \rangle = \langle \epsilon_{2}^{2} \rangle$ is the
variance of the intrinsic source ellipticities.  For our sample,
$\sigma_{\epsilon}\simeq 0.44$ and $g\simeq 1.61$. While the small
source density in our sample prohibits a mapping of the shear, the
lensing effect can be studied statistically. This can be achieved by
measuring the shear correlation functions $C_{1}(\theta)=g^{-2}\langle
\epsilon^{r}_{1}(0)\epsilon^{r}_{1}(\theta) \rangle$ and
$C_{2}(\theta)=g^{-2}\langle
\epsilon^{r}_{2}(0)\epsilon^{r}_{2}(\theta) \rangle$ , where the
average is over source pairs with separation $\theta$, and the rotated
ellipticities $\epsilon^{r}$ are measured along the separation vector.

\section{Theory and Sensitivity}
To estimate the expected lensing signal, we consider the four CDM
models listed in Table~1 \cite{kam98}. While model 1 is
COBE-normalized, models 2-4 are essentially cluster-normalized
($\sigma_{8} \Omega^{0.53} = 0.6 \pm 0.1$, see \cite{via96}) and are
thus more realistic. For each model, we list the {\it rms} shear
$\sigma_{\gamma}(1^{\circ})$ in square cells of size
$\theta_{c}=1^{\circ}$, along with the signal-to-noise ratio
SNR$(1^{\circ}) \simeq 2^{-1} g^{2} \sigma_{\gamma}^{2}(1^{\circ})
\sigma_{\epsilon}^{-2} n A^{\frac{1}{2}} \theta_{c}$ for detecting
this excess variance with FIRST.  For the cluster-normalized models,
an {\it rms} shear of about 1.5\% is expected to be detectable at the
$\sim 4\sigma$ level, in this fashion.

\begin{center}
\begin{tabular}{rcccccccc}
\multicolumn{9}{l}{Table 1: Weak lensing signal expected for several
CDM models with the}\\
\multicolumn{9}{l}{FIRST survey (7,200 deg$^{2}$)}\\
\hline
\\
\multicolumn{1}{c}{model}&
\multicolumn{1}{c}{$\Omega$}&
\multicolumn{1}{c}{$h$}&
\multicolumn{1}{c}{$n$}&
\multicolumn{1}{c}{$\Omega_{b}h^{2}$}&
\multicolumn{1}{c}{$\sigma_{8}\Omega_{0}^{0.53}$}&
\multicolumn{1}{c}{$\sigma_{\gamma}(1^{\circ})$} &
\multicolumn{1}{c}{SNR$(1^{\circ})$}&
\multicolumn{1}{c}{SNR$({\cal L})$}\\
\hline
\\
1  & 1 & 0.50 & 1 & 0.012 & 1.21 & 0.022 & 9.9 & 31\\
2 & 1 & 0.50 & 0.8 & 0.025 & 0.71 & 0.014 & 4.0 & 10\\
3 & 0.4 & 0.65 & 1 & 0.015 & 0.65 & 0.013 & 3.5 & 8\\
4 & 1 & 0.35 & 1 & 0.015 & 0.74 & 0.015 & 4.6 & 11\\
\\
\hline
\end{tabular}
\end{center}
\smallskip

The shear correlation functions expected for each model are shown in
Figure~1.  Note that our linear predictions would be revised upward
for $\theta \lesssim 10'$ if nonlinear evolution were included
\cite{jai97}.  In the absence of lensing, the {\it rms} fluctuation of
$C_{i}(\theta)$ averaged in an interval of width $\Delta \theta$,
would be $\sigma \left[ C_{i}(\theta) \right] = g^{-2}
\sigma_{\epsilon}^{2} N_{\rm pairs}^{-\frac{1}{2}}(\theta)$, where
$N_{\rm pairs}(\theta)\simeq 2 \pi (\ln 10) n^{2} A \theta^{2} \Delta
\log \theta$ is the number of source pairs in this interval. As the
straight lines in Figure~1 demonstrate, we will be able to detect a
signal for $C_{1}(\theta)$ and $C_{2}(\theta)$ for $\theta$ ranging
from 0.2 to 20 degrees. The complete lensing signal can be measured
by performing a maximum likelihood analysis for the power spectrum
amplitude \cite{kam98}.  The signal-to-noise ratio, SNR$({\cal
L})$, expected with this method is listed in the last column of
table~1 and is about 10 for cluster-normalized models.

\begin{figure}
\centerline{\vbox{
\psfig{figure=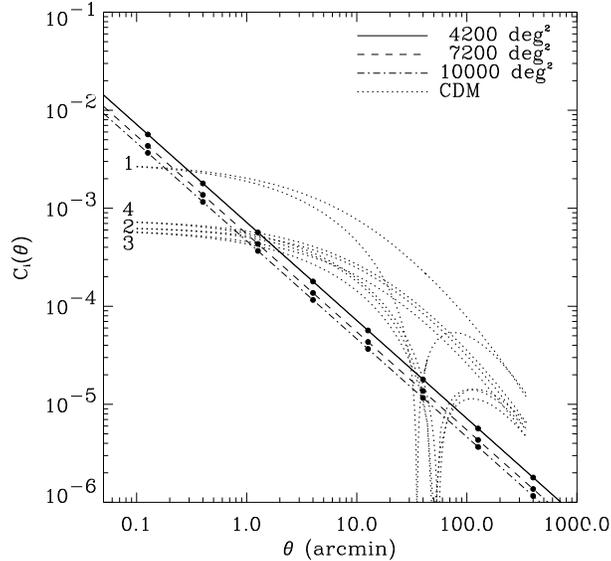,height=8.cm}
}}
\caption[]{Sensitivity for measurement of the shear correlation
functions $C_{i}$ with the FIRST survey. The dotted curves show the
expected values of $C_{1}(\theta)$ (top) and $C_{2}(\theta)$ (bottom),
for each of the four CDM models of Table~1, as labeled on the
left-hand side.  The straight lines show the $1\sigma$ sensitivity for
the present, allocated, and nominal sky coverage of the FIRST survey,
binned in $\Delta \log \theta = 0.5$ intervals.}
\end{figure}

\section{Systematic Effects}
Because the weak-lensing signal is only of the order of 1\%,
systematic effects must be carefully accounted for. The following
briefly describes the major effects which can introduce
spurious ellipticity correlations.

\smallskip
\noindent {\bf Source Fragmentation}: A significant fraction of radio
sources have a double-lobe structure, and are broken into two
components by the object finder.  Because these fragments tend to be
aligned, this produces ellipticity correlations on small angular
scales. An analysis of the excess signal in the correlation function
of FIRST source positions reveals, however, that this effect is
negligible for $\theta \gtrsim 10'$.

\smallskip
\noindent {\bf Correlated Noise:} Because the FIRST survey was derived
from interferometric observations, the noise in its images is
spatially correlated, as is apparent in the presence of ``stripes'' in
the noise. The consequence of noise correlation on source
ellipticities was investigated in detais in ref.~\cite{ref98b}, where
it was found that, while the effect is important on small scales, it
is negligible for $\theta \gtrsim 1'$.

\smallskip
\noindent {\bf Beam Distortion:} Spurious shape correlations can also
be produced by spatial variations of the effective convolution beam.
The beam shape mostly depends on the orientations of the VLA antennae
at the time of the observation, and can thus be modeled and corrected
for, a posteriori. We are currently studying the magnitude of this
effect and its consequence for weak lensing measurements.

\section{Conclusions}
If, as we expect, the effect beam distortion can be corrected for, the
FIRST survey will allow us to detect, or at least to set a tight upper
limit on, the weak lensing power spectrum on 0.2-20 degree scales.
Since these scales correspond to the linear regime, our measurement
will be easy to compare to theoretical predictions. Our preliminary
measurement of the shear correlation function already leads to upper
limits that are close to the cluster-normalized CDM predictions. Our
experiment will complement similar measurements in the optical band,
which are mostly sensitive to smaller angular scales.

\acknowledgements{We thank Nick Kaiser, Frazer Owen, Jacqueline van
Gorkom, David Schiminovich, and Don Neill for useful discussions.
This work was supported in part by the NASA MAP/Midex program, the
A.P. Sloan Foundation, grant DE-FG02-92ER40699, and NASA grant
NAG5-3091.}


%
\vfill
\end{document}